\documentclass[lettersize,journal]{IEEEtran}
\usepackage{amsmath,amsfonts}
\usepackage{algorithmic}
\usepackage{algorithm}
\usepackage{array}
\usepackage[caption=false,font=normalsize,labelfont=sf,textfont=sf]{subfig}
\usepackage{textcomp}
\usepackage{stfloats}
\usepackage{url}
\usepackage{verbatim}
\usepackage{graphicx}
\usepackage{cite}
\usepackage[capitalize,noabbrev]{cleveref}
\usepackage{wrapfig}
\usepackage{pifont}
\usepackage{amsfonts}  % blackboard math symbols
\usepackage{nicefrac}  % compact symbols for 1/2, etc.
\usepackage{xcolor} % colors
\usepackage{multirow}
\usepackage{multicol}
\usepackage{bbding}
\usepackage{booktabs} % 支持双线表格

\begin{document}
	
	\title{GENE-FL: Gene-Driven Parameter-Efficient Dynamic Federated Learning}
	\author{Shunxin~Guo, Jiaqi~Lv, Qiufeng~Wang,~and~Xin Geng% <-this % stops a space
%		\thanks{This research was supported by the National Science Foundation of China (62125602, 62076063, 62302093). (\textit{Corresponding authors: Hongsong Wang; Xin Geng}).}% <-this % stops a space
%		\thanks{
%			The authors are with the School of Computer Science and Engineering, Southeast University, Nanjing 211189, China, and also with the Key Laboratory of New Generation Artificial Intelligence Technology and Its Interdisciplinary Applications (Southeast University), Ministry of Education, China. (e-mail: sxguo@seu.edu.cn; hongsongwang@seu.edu.cn; xgeng@seu.edu.cn).}
%	}
%	
%	\author{IEEE Publication Technology,~\IEEEmembership{Staff,~IEEE,}
%		% <-this % stops a space
%\thanks{This paper was produced by the IEEE Publication Technology Group. They are in Piscataway, NJ.}% <-this % stops a space
\thanks{Manuscript received April 19, 2021; revised August 16, 2021.}}
	\markboth{Journal of \LaTeX\ Class Files,~Vol.~14, No.~8, August~2021}%
	{Shell \MakeLowercase{\textit{et al.}}: A Sample Article Using IEEEtran.cls for IEEE Journals}
	
%	\IEEEpubid{0000--0000/00\$00.00~\copyright~2021 IEEE}
	
	\maketitle
	
	\begin{abstract}
		Real-world \underline{F}ederated \underline{L}earning systems often encounter \underline{D}ynamic clients with \underline{A}gnostic and highly heterogeneous data distributions (DAFL), which pose challenges for efficient communication and model initialization. To address these challenges, we draw inspiration from the recently proposed Learngene paradigm, which compresses the large-scale model into lightweight, cross-task meta-information fragments. Learngene effectively encapsulates and communicates core knowledge, making it particularly well-suited for DAFL, where dynamic client participation requires communication efficiency and rapid adaptation to new data distributions. Based on this insight, we propose a Gene-driven parameter-efficient dynamic Federated Learning (GENE-FL) framework. First, local models perform quadratic constraints based on parameters with high Fisher values in the global model, as these parameters are considered to encapsulate generalizable knowledge. Second, we apply the strategy of parameter sensitivity analysis in local model parameters to condense the \textit{learnGene} for interaction. Finally, the server aggregates these small-scale trained \textit{learnGene}s into a robust \textit{learnGene} with cross-task generalization capability, facilitating the rapid initialization of dynamic agnostic client models. Extensive experimental results demonstrate that GENE-FL reduces \textbf{4 $\times$} communication costs compared to FEDAVG and effectively initializes agnostic client models with only about \textbf{9.04} MB.
	\end{abstract}
	
	
	\section{Introduction}
	\IEEEPARstart{F}{ederated} Learning (FL)~\cite{mcmahan2017communication} has shown great promise in the field of distributed learning across devices, allowing multiple clients to collaboratively train a shared global model without exposing private data~\cite{chen2022calfat,qi2025cross}. The integration of FL enhances the safety and efficiency of real-world AI applications in various domains, including medical impact analysis~\cite{ng2021federated,guan2024federated}, personalized recommendation systems~\cite{wu2023personalized,imran2023refrs}, and intelligent transportation systems~\cite{shinde2023joint,pandya2023federated}.
	
	In real-world applications, FL is increasingly being deployed in dynamic and diverse environments, where participating clients often exhibit agnostic classes and heterogeneous data distributions, forming a Dynamic Agnostic Federated Learning (DAFL) scenario. This introduces two critical challenges that must be addressed simultaneously: \textbf{\ding{182}Efficient Communication}: The limited bandwidth and high communication costs in FL systems can significantly impact overall performance and scalability. \textbf{\ding{183}Rapid Adaptation to Agnostic Client Models}: Dynamically joining clients may have agnostic data distributions, requiring effective model initialization to avoid expensive retraining and slow convergence.
	Common approaches to improve communication efficiency include model pruning and compression~\cite{karimireddy2020scaffold,haddadpour2021federated}, one-shot FL~\cite{jhunjhunwala2024fedfisher,elmahallawy2023one,zhang2022dense,andrewone}, reducing local updates to control communication costs~\cite{karimireddy2020scaffold}, and using class prototypes to minimize communication overhead~\cite{tan2022fedproto,zhang2024fedtgp}. 
	While these approaches are well-suited for FL scenarios with a stable number of clients and class distributions, they exhibit limitations in the DAFL setting. Specifically, pruned models or trained class prototypes are based only on information from fixed training data, making it difficult to generalize to agnostic clients with new data distributions or classes that were not included during the pruning process. This highlights the core goal of DAFL: \textit{How can we design a framework that enables efficient communication between server and clients while ensuring effective initialization of agnostic client models?}
	
	\begin{figure*}[ht]
		\begin{center}
			\includegraphics[width=0.9\linewidth]{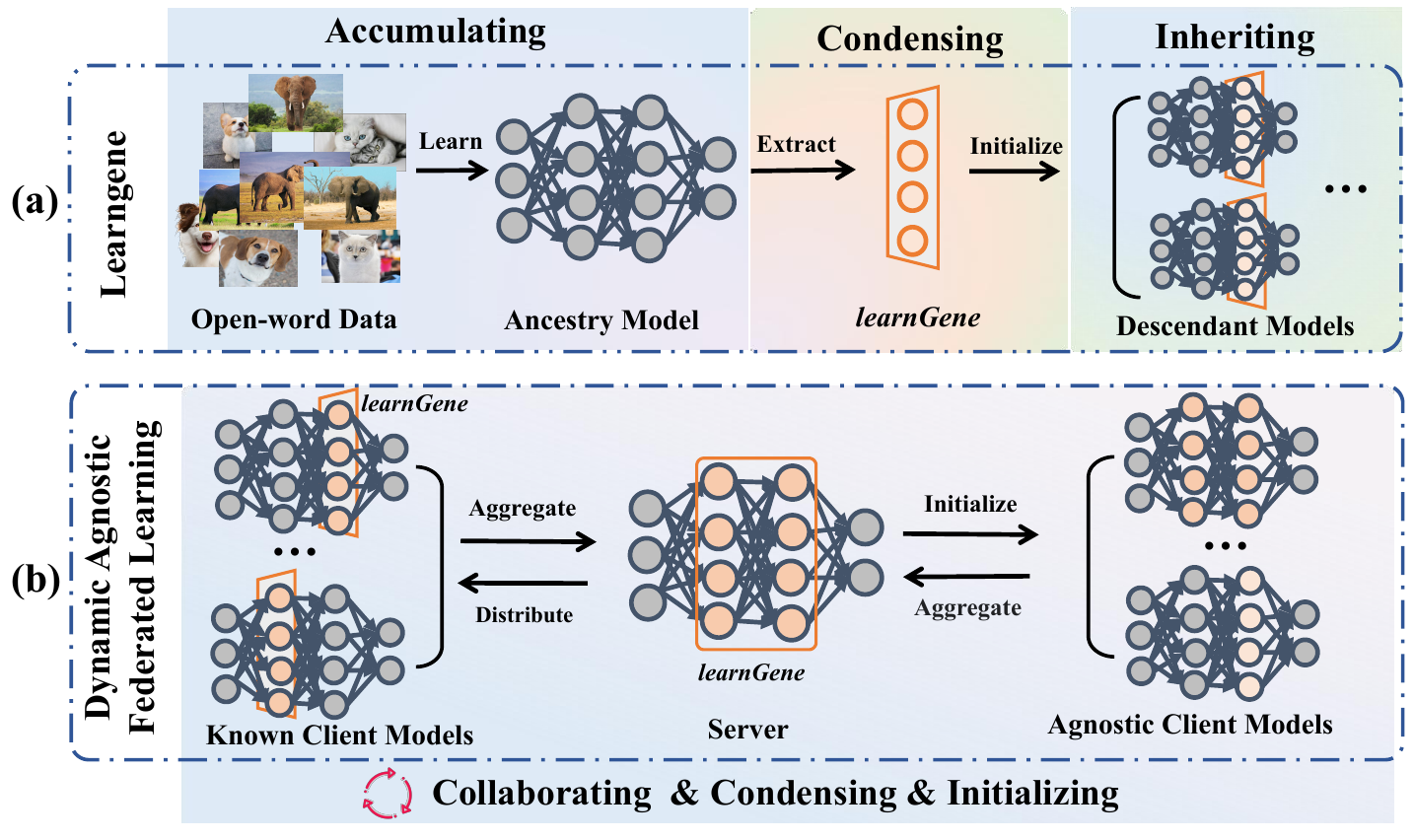}
			\caption{Illustration of Dynamic Agnostic Federated Learning and Learngene. In the accumulating, condensing, and inheriting processes of the Learngene, dynamic agnostic federated learning can achieve a corresponding organic integration.}
			\label{intro}
		\end{center}
	\end{figure*}
	
	To achieve this goal, we are inspired by a novel Learngene paradigm~\cite{wang2022learngene, wang2023learngene} that involves learning from open-world data to accumulate knowledge for training a large-scale ancestry model. This model is then condensed into a lightweight \textit{learnGene}~\footnote{``Learngene'' refers to the learning framework, while ``\textit{learnGene}'' denotes the condensed information piece of the model.}, which encapsulates shared meta-knowledge and is inheritable. Subsequently, models for different tasks inherit \textit{learnGene} for rapid and effective initialization, as shown in Figure~\ref{intro} (a). The ability of Learngene to efficiently encapsulate and transfer core knowledge makes it particularly well-suited for DAFL, where dynamic client participation demands both communication efficiency and rapid adaptation to new data distributions.
	
	Building on this insight, we propose a Gene-driven parameter-efficient dynamic Federated Learning framework (GENE-FL) tailored for DAFL scenarios, with the summarized process illustrated in Figure~\ref{intro}(b): \textbf{(\romannumeral1) Condensing}: the known client models in the FL system are condensed into small-scale \textit{learnGene}s through smooth updates. \textbf{(\romannumeral2) Collaborating}: clients and the server engage in efficient communication based on \textit{learnGene}s, where the server collaborates to aggregate the \textit{learnGene}s uploaded by participated clients. \textbf{(\romannumeral3) Initializing}: the agnostic clients dynamically join the FL system and utilize the \textit{learnGene}s downloaded from the server to quickly initialize their models. The three processes in the framework complement each other and can be executed in parallel, enabling efficient communication and scalability.
	Specifically, to address the communication latency issue commonly encountered in FL systems (where the slowest participating client affects communication time when using a single global model)~\cite{vahidian2023efficient}, we introduce a one-shot clustering method to obtain multiple cluster models. Initially, each local model is smoothly updated based on its corresponding cluster model, absorbing the knowledge from other participants. To ensure the generalization of the condensed \textit{learnGene} across data distributions, we impose quadratic constraints on the local model using the elastic \textit{learnGene} that is obtained by partitioning the cluster model based on Fisher information. Second, each client performs sensitivity analysis on parameter updates to extract \textit{learnGene}, facilitating \textbf{\ding{182}} efficient communication with the server. Finally, the server dynamically updates the cluster \textit{learnGene} based on the uploaded \textit{learnGene}s, enabling \textbf{\ding{183}} rapid adaptation to agnostic client models. 
	
	In summary, we analyze the adaptability of the Learngene framework to the goal of DAFL, and based on this, propose the first method that leverages Learngene technology to simultaneously address the dual challenges.
	
\section{Related Work}\label{sec:rw}
\textbf{Federated Learning.} Federated Learning (FL) is a distributed learning framework designed to safeguard user data privacy~\cite{mcmahan2017communication,qi2023cross,Wang2023DaFKD}. Compared to centralized learning, FL faces several unique challenges, including non-independently identically distributed and imbalanced data, as well as limited communication bandwidth~\cite{mu2021fedproc,2020Briggs,lee2022preservation,lifedcompass}. An intuitive approach to reduce communication costs is to quantify weights directly and upload them sparsely~\cite{yi2024fedpe,jiang2023computation,jiang2023complement,shi2024efficient}. The FedDrop~\cite{caldas2018expanding} reduces the computational burden of local training and the corresponding communication costs in FL. The transferable model design in FEDLPS~\cite{jia2024fedlps} uses an adaptive channel model pruning algorithm. Efforts have also been directed towards one-shot FL~\cite{jhunjhunwala2024fedfisher,elmahallawy2023one}, which aims to achieve satisfactory models with only one round of communication, but requires the shared dataset. 
Additionally, using class prototypes for low-cost communication has gained attention~\cite{tan2022fedproto}. Further, FedTGP~\cite{zhang2024fedtgp} proposes using adaptive margin-enhanced contrastive learning to train global prototypes on the server.
However, most existing studies focus on FL scenarios where the number of participating clients and the number of classes remain stable, with limited attention given to the critical issue of model initialization and optimization when new classes or clients are dynamically introduced. In contrast, we propose a Gene-driven dynamic agnostic FL framework, which leverages collaborative learning among existing clients and condensing \textit{learnGene} to reduce communication costs, and uses \textit{learnGene} to initialize the agnostic client models.

%we propose a dynamic agnostic federated learning with Learngene framework, which condenses lightweight \textit{learnGene} for participating client model to reduce unnecessary resource consumption in dynamic scenarios.

\textbf{Learngene.} The Learngene~\cite{lin2024linearly,xiainitializing,xia2024transformer,wangvision,wangcluster,li2024facilitating,feng2024wave,xie2024kind}, as a novel paradigm based on the inheritance principles from biology, enables the condensation of a large-scale ancestral model into \textit{learnGene} to adaptively initialize models for various descendant tasks.
Wang et al.~\cite{wang2022learngene} first proposed Learngene based on gradient information from the ancestral model, using limited samples to initialize descendant models. 
Furthermore, they summarized the three processes of Learngene~\cite{wang2023learngene}: accumulating, condensing, and inheriting.
To facilitate the rapid construction of numerous networks with different complexity and performance trade-offs, a \textit{learnGene}~\cite{shi2024building} pool is proposed to satisfy low-resource constraints. Simultaneously, GTL~\cite{feng2024transferring} demonstrated that the transfer of core knowledge through \textit{learnGene} can be both sufficient and effective for neural networks.
These mentioned approaches underscore the promise of the Learngene paradigm and its feasibility in reducing costs while preserving the essential knowledge of models. Based on this insight, we propose a gene-driven FL framework for DAFL.

\textbf{Fisher Information Matrix.}
The Fisher Information Matrix (FIM)~\cite{barrett1995objective,ly2017tutorial} is a key concept in statistical estimation theory that encapsulates the information that unknown parameters hold about a random distribution. In deep learning, the FIM has been used to study adversarial attacks~\cite{zhao2019adversarial}, guide optimization, and evaluate the information content of parameters~\cite{fasina2023neural,jhunjhunwala2023towards,vallisneri2008use}. For example,~\cite{zhao2019adversarial} utilizes the eigenvalues of FIM derived from a neural network as features and trains an auxiliary classifier to detect adversarial attacks on the eigenvalues. The layer-wise correlation propagation method~\cite{binder2016layer} uses the diagonal of FIM to quantify the importance of parameters, thereby improving the interpretability of the model. The Elastic Weight Removal method~\cite{daheim2023elastic} weights the individual importance of the parameters via FIM to eliminate hallucinations. These methods all use the diagonal approximation of the FIM to reduce computational complexity and promote a more efficient learning process based on the Fisher information of the parameters.

\section{Methodology}
\textbf{Notions.} Let $\mathcal{N}$ be the set of known clients with the size of $N$, where the non-iid distributed training data on $i$-th client is denoted as $\mathcal{D}_{i}=\left\{\left(x_{i}, y_{i}\right)\right\}$, $i \in \mathcal{N}, x_{i}, y_{i}$ are the corresponding data pair. Similarly, $\mathcal{M}$ denotes the set of agnostic clients with the size of $M$. Additionally, we aim to group clients with similar data distributions, such that clients within the same cluster can leverage each other's data for improved performance in federated learning. On the server side, the known clients $\mathcal{N}$ that have already participated in training are grouped into $K$ clusters (denoted as $k$) based on the distributional similarity between their data subspaces, using a one-shot clustering approach as detailed in \cite{vahidian2023efficient}. Therefore, the server contains $K$ cluster models, where a client $i$ belonging to cluster $k$ has a parameterized classification network $\theta_{k,i}$, and the corresponding cluster model is $\Theta_k$. 
Generally, each client $i$ optimizes its model by minimizing the classification loss, as follows:
\begin{equation}
	\label{eq:cls}
	\mathcal{L}_{cls}=\mathbb{E}_{x_i, y_i \sim \mathcal{D}_{i}} \Phi \left(\left(x_i \mid y_i ; \theta_{k,i}\right), y_i\right),
\end{equation}
where $\Phi$ is the Cross-Entropy loss function and $y_i$ is the ground truth label.

\textbf{Definition 1} (Agnostic and Heterogeneous Data Distributions). \textit{Let \( p_i(y) \) and \( p_j(y) \) denote the label distributions for clients \( i \) and \( j \), respectively, and \( \mathcal{Y}_n \) and \( \mathcal{Y}_m \) represent the set of known and agnostic classes, respectively. Under DAFL, the label distribution among the known classes \( \mathcal{Y}_n \) may vary across different known clients, but the class-conditional distributions \( p(x|y) \) for these classes remain invariant across clients. Additionally, the agnostic clients introduce novel classes \( \mathcal{Y}_m \) that do not overlap with the known classes, i.e.,}
\begin{multline}
	p_i(y) \neq p_j(y) \quad \forall y \in \mathcal{Y}_n \\
	\text{but} \quad p_i(x|y) = p_j(x|y); \\
	\text{and} \quad \mathcal{Y}_n \cap \mathcal{Y}_m = \emptyset.
\end{multline}

\begin{figure*}[ht]
	\begin{center}
		\includegraphics[width=0.85\linewidth]{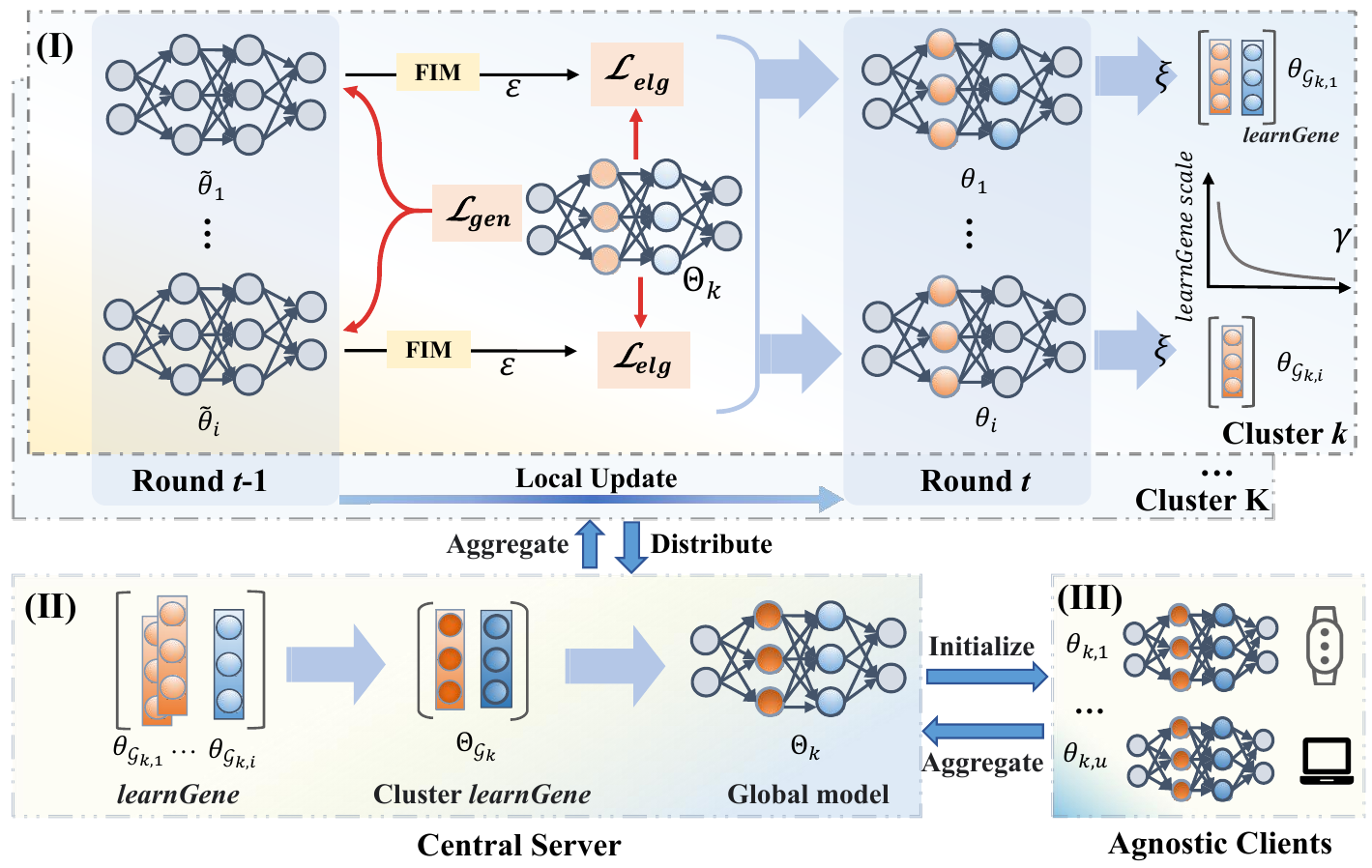}
		\caption{Illustration of training process of GENE-FL, which includes (\uppercase\expandafter{\romannumeral1}) \textit{learnGene} Condensation via Smooth Updating, (\uppercase\expandafter{\romannumeral2}) \textit{learnGene} Collaborative Aggregation, and (\uppercase\expandafter{\romannumeral3}) \textit{learnGene} Initial Agnostic Client Model.}
		\label{method}
	\end{center}
\end{figure*}

\textbf{Method Overview.}	
The proposed GENE-FL framework consists of three learning processes, as shown in Figure~\ref{method}.
During the $t$-th round, the local models perform smooth updates based on the cluster model and execute quadratic regularization using the elastic \textit{learnGene} partitioned by the Fisher Information Matrix (FIM) to improve the adaptability of the local models to the client data distribution.
The optimal \textit{learnGene} is identified based on the layer-wise sensitivity score $\xi$ with the previous model $\tilde{{\theta}}_{k,i}$.
Participating clients then upload their individual \textit{learnGene} ($\theta_{\mathcal{G}_{k,1}}, \cdots,\theta_{\mathcal{G}_{k,i}}$) to the server for dynamic aggregation and subsequent distribution to them.
When the agnostic client makes a model request, the server sends the nearest cluster \textit{learnGene} to facilitate the initialization of its model.

%\begin{figure*}[!htb]
%	\centering
%	\includegraphics[width=0.75\linewidth]{images/method.pdf}
%	\caption{Illustration of training process of GENE-FL, which includes (\uppercase\expandafter{\romannumeral1}) Learngene Smooth Updating, (\uppercase\expandafter{\romannumeral2}) Learngene Dynamic Aggregation, and (\uppercase\expandafter{\romannumeral3}) Learngene Initial Agnostic Model.}
%	\vspace{-4mm}
%	\label{method}
%\end{figure*} 

\subsection{\textit{learnGene} Condensation via Smooth Updating}
During local updates, the goal is to compress a unique \textit{learnGene} for each client, thereby reducing communication costs interacting with the server. In particular, local models approximate the corresponding cluster model constraints for smooth updates and apply quadratic regularization on the elastic \textit{learnGene} partitioned by FIM to effectively capture generalizable knowledge.

\textbf{Local Smooth Updating Based on Cluster Model.}
We leverage the cluster model obtained after collaborative learning to impose smooth constraints on the local models.
These models trained locally exhibit similarity within the same cluster, uploading their itself lightweight \textit{learnGene} for client-server interaction, effectively mitigating communication costs. Traditional FL methods typically apply a uniform regularization strength to all parameters of the model, ignoring inherent differences in magnitudes among parameters, leading to biased selection of the \textit{learnGene}. To address this, we design an adaptive smoothness constraint for each layer and determine the strength of the constraint based on the size of its parameters. The local update of the $i$-th model is expressed as:
\begin{equation}
	\label{lp}
	\theta_{k,i} \leftarrow \tilde{{\theta}}_{k,i} - \alpha \nabla_{\theta} \mathcal{L}_{gen}(\tilde{{\theta}}_{k,i}, \mathcal{D}_{i}),
\end{equation}
where $\mathcal{D}_{i}$ represents the private dataset of client $i$, $\alpha$ is the learning rate, $\tilde{{\theta}}_{k,i}$ is the previous model, and $\nabla_{\theta}\mathcal{L}_{gen}$ is the gradient of the first smooth loss function. This is calculated as the sum of the Cross-Entropy loss and the L2 norm of the difference between the current model $\theta_{{k,i}}$ and the cluster model $\Theta_{{k}}$:
\begin{equation}
	\label{eq:gen}
	\mathcal{L}_{gen}= \left\|\theta_{{k,i}}-\Theta_{{k}}\right\|_{2}.
\end{equation}

\textbf{Quadratic Local Smooth Updating Based on Elastic \textit{learnGene}.}
Considering the effective initialization of agnostic client models during the learning process, lightweight \textit{learnGene} should have high informational content to effectively predict untrained classes.
Fisher information matrix (FIM) quantification of model parameters provides rich information content~\cite{jhunjhunwala2024fedfisher}.
Inspired this, we introduce the FIM that diagonal is used to weight the importance of each parameter of the model and determine the penalty size for changing the parameter in client training. 
We can obtain a good approximation to the diagonal of the Fisher values for each parameter indexed by $j$ in the model $\tilde{\theta}_{i}$ (refers to $\tilde{\theta}_{k,i}$), as follows:
\begin{equation}
	\label{eq:fisher}
	F_{i,j}=\mathbb{E}\left[\left(\frac{\partial \log h( \tilde{\theta}_i \mid \mathcal{D}_i)}{\partial \tilde{\theta}_{i,j}}\right)^{2}\right],
\end{equation}
where $h( \tilde{\theta}_i \mid \mathcal{D}_i)$ is the likelihood function that represents the fitness of the model parameters under the data $\mathcal{D}_i$.
% Based on this, we can obtain the elastic \textit{learnGene} from the training model.

The elastic \textit{learnGene} we seek is a model fragment that reflects the shared knowledge of the sample data, selected based on Fisher values computed from the model and private data. Specifically, if the Fisher value is below a given threshold $\varepsilon$, the model parameters with small local model update variations (i.e., its own \textit{learnGene}) are retained, while the remaining parameters are updated using the aggregated consensus model (knowledge shared across clients):
\begin{equation}
	\label{eq:sec}
	\theta'_{k, i} = \left\{\begin{array}{ll}
		\tilde{\theta}_{k,i,j}, & \hat{F}_{i, j} \le  \varepsilon  \\
		\Theta_{k,j}, &  { otherwise }, 
	\end{array}\right.
\end{equation}
where $\hat{F}_{i, j}=\frac{F_{i, j}-\min \left(F_{i}\right)}{\max \left(F_{i}\right)-\min \left(F_{i}\right)}$ represents the normalization of $F_{i, j}$. 

The local model is restructured according to the FIM into a combination of the \textit{learnGene} and the cluster model, aiming to approximate its own \textit{learnGene} to the cluster elastic \textit{learnGene}. 
This enables learning of shared knowledge within the same cluster to reduce redundant parameters, thereby lowering communication costs. 
Then the quadratic regularization update of the current local model based on the cluster elasticity \textit{learnGene} is expressed as:
\begin{equation}
	\label{elg}
	\theta_{k,i} \leftarrow \theta_{k,i} - \alpha \nabla_{\theta} \mathcal{L}_{elg}(\theta'_{k,i}, \mathcal{D}_{i}),
\end{equation}
where $\nabla\mathcal{L}_{elg}$ is the gradient of the elasticity loss $\mathcal{L}_{elg}$, calculated as the L2 norm of the difference between the local model’s \textit{learnGene} and the cluster’s elastic \textit{learnGene}:
\begin{equation}
	\label{eq:elg}
	\mathcal{L}_{elg}=\left\|\theta'_{k,i}-\Theta_k\right\|_{2}.
\end{equation}
In summary, the total optimization objective for local updates is defined as $\mathcal{L}_{all} = \mathcal{L}_{cls} + \lambda_1\mathcal{L}_{gen} + \lambda_2\mathcal{L}_{elg}$, where $\mathcal{L}_{cls}$ represents the classification loss, and $\lambda_1$ and $\lambda_2$ are hyperparameters that regulate $\mathcal{L}_{gen}$ and $\mathcal{L}_{elg}$, respectively.

\textbf{Condensation of \textit{learnGene}.} 
After training the local model based on private data, we precisely identify each client's \textit{learnGene}, enabling the acquisition of meta-knowledge about the model. The strategy is determining the contribution of each layer, guided by the parameter changes observed after model training.
The score of the $l$-th layer $\xi_{k,i}^{(l)}$ can be calculated on the locally trained model $\theta_{k,i}$ and previous model $\tilde{\theta}_{k,i}$, as follows: 
\begin{equation}
	\label{eq:cos}
	\xi_{k,i}^{(l)}=\frac{cos\left({\theta}^{(l)}_{k,i}, \tilde{{\theta}}^{(l)}_{k,i }\right)}{{dim}\left(\theta^{(l)}_{k,i}\right)},
\end{equation}
where $dim(\cdot)$ represents the number of parameters in the $l$-th layer, normalized as $\sum_{l=1}^{L} \xi_{k,i}^{(l)} = 1$ to enable comparisons at a unified granularity. $cos(\cdot)$ denotes cosine similarity, which can be substituted with alternative measures such as L1, L2, or Earth Mover Distance. The term $\xi_{k,i}^{(l)}$ quantifies the difference between ${\theta}^{(l)}_{k,i}$ and $\tilde{{\theta}}^{(l)}_{k,i}$, capturing the degree of updates to the local model driven by private data. Intuitively, higher $\xi_{k,i}^{(l)}$ values indicate a more pronounced influence of private data on the local model, reflecting its personalized components. Conversely, lower $\xi_{k,i}^{(l)}$ values correspond to layers that primarily represent generalized information less sensitive to private data. This generalized information is beneficial for model initialization in new tasks, consistent with what we seek for \textit{learnGene}.

The mask set $S_{k,i}$ is to sort the $\xi_{k,i}$ values of the $L$ layer in descending order, set the top $\gamma$ layer to 1, and the others to 0, to get the sorted model $\theta_{\mathcal{G}_{k,i}^{(l)}} =\theta_{k,i} \odot S_{k,i}$.
The scale of \textit{learnGene} progressively decreases during the model training and optimization process until it stabilizes at a threshold $\gamma$, determined by a performance-based adaptive training procedure. This gradually reduces the communication cost between the client and the server, and while ensuring the performance of the model, it also mines the generalization \textit{learnGene} that can initialize the new client model. Algorithm~\ref{alg:dflg} describes the process of local model updates and the condensation of the \textit{learnGene}.
\begin{algorithm}[!htb]
	\caption{GENE-FL: Updating and Condensation}
	\label{alg:dflg}
	\begin{algorithmic}[1]
		\STATE{\bfseries Input:}
		Local update epochs $E_l$, $N_k$ participants in the $k$-th cluster, private data of the $i$-th client $\mathcal{D}_i$, cluster model $\Theta_k$, local model $\theta_{k,i}$ and previous local model $\tilde{\theta}_{k,i}$, learning rate $\alpha$.
		\STATE{\bfseries Output:} Local \textit{learnGene} $\{\theta_{\mathcal{G}_{k,i}}\}_{i=1}^{N_k}$.
		\FOR{$i = 1,2,\cdots, N_k$ \textbf{in parallel}}
		\STATE Receive $\Theta_k$ from server
		\STATE Compute $F_{k,i} \leftarrow (\mathcal{D}_{i}, \tilde{\theta}_{k,i})$ using Eq.~(\ref{eq:fisher})
		\FOR{$e = 1,2,\cdots, E_l$}
		\STATE Update $\theta_{k,i} \leftarrow \tilde{\theta}_{k,i} - \alpha \nabla_{\theta} \mathcal{L}_{gen}(\tilde{\theta}_{k,i}, \mathcal{D}_{i})$ using $\mathcal{L}_{gen}$ from Eq.~(\ref{eq:gen})
		\STATE Compute $\theta'_{k,i} \leftarrow (F_{k,i}, \Theta_k, \tilde{\theta}_{k,i}, \varepsilon)$ using Eq.~(\ref{eq:sec})
		\STATE Update $\theta_{k,i} \leftarrow \theta_{k,i} - \alpha \nabla_{\theta} \mathcal{L}_{elg}(\theta'_{k,i}, \mathcal{D}_{i})$ using $\mathcal{L}_{elg}$ from Eq.~(\ref{eq:elg})
		\ENDFOR
		\STATE Compute $\xi_{k,i} \leftarrow (\tilde{\theta}_{k,i}, \theta_{k,i})$ using Eq.~(\ref{eq:cos})
		\STATE Obtain mask set $S_{k,i}$ by sorting $\xi_{k,i}$ of the $L$ layers
		\STATE Compute $\theta_{\mathcal{G}_{k,i}} \leftarrow \theta_{k,i} \odot S_{k,i}$
		\ENDFOR
	\end{algorithmic}
\end{algorithm}

\subsection{\textit{learnGene} Collaborative Aggregation}
In the server, our goal is to maintain a unified cluster \textit{learnGene} for each cluster, which encapsulates the generalization parameter information of all relevant local models within the cluster, allowing effective initialization of newly agnostic client models.
The \textit{learnGene} layers that are common to all participants are aggregated to obtain clusters \textit{learnGene}, while the others retain the previous cluster model.
The formula for aggregating the cluster \textit{learnGene} is as follows:
\begin{equation}
	\label{eq:agg1}
	\Theta_{\mathcal{G}_k^{(l)}} = \frac{1}{N^{(l)}_k} \sum_{i=1}^{N^{(l)}_k} \theta_{\mathcal{G}_{k,i}^{(l)}}, 
\end{equation}
where $\Theta_{\mathcal{G}_k^{(l)}}$ represents the parameters of the $l$-th layer within the $k$-th cluster \textit{learnGene}. Additionally, $N^{(l)}_k$ denotes the number of client \textit{learnGene} that encompass the $l$-th layer, while $\theta_{\mathcal{G}_{k,i}^{(l)}}$ signifies the parameters of the $l$-th layer within the $i$-th \textit{learnGene} belonging to the $k$-th cluster. The updated cluster model is denoted as the aggregated \textit{learnGene} and partial previous cluster model: $\Theta_k = [\Theta_{\mathcal{G}_k} ; \tilde{\Theta}_k]$.

\subsection{\textit{learnGene} Initial Agnostic Model}
In the dynamic and agnostic FL scenario, when a new client $i$ joins, we recommend applying truncated singular value decomposition~\cite{li2024beyond} to its private data sample $\mathcal{X}_i$ to obtain the components that describe the underlying data distribution. The decomposition is defined as:
\begin{equation}
	\label{eq:svd}
	\mathcal{X}_{i,d} = \mathbf{U}_{i,d} \mathbf{\Sigma}_{i,d} \mathbf{V}_{i,d}^T,
\end{equation}
where $\mathbf{U}_{i,d} = [\mathbf{u}_1, \mathbf{u}_2, \ldots, \mathbf{u}_d]\in \mathbb{R}^{m \times d}$ (with $d \ll \text{rank}(\mathcal{X}_i)$ and $m$ denotes the number of samples for client $i$) represents the top $d$ most significant left singular vectors, capturing the essential features of the underlying data distribution.
We follow the~\cite{vahidian2023efficient} and select $d = 5$ to mitigate the risk of data leakage. Additionally, to facilitate linear algebraic computations, we transform the matrix $\mathbf{U}_{i,d}$ into a vector $\mathbf{u}_{i,d} \in \mathbb{R}^{md \times 1}$.

Then, client $i$ upload $\mathbf{u}_{i,d}$ to the server for requesting \textit{learnGene}. Let $\mathbf{u}_k$ be the mean vector of the $k$-th cluster. The server calculates the distance $d_{i,k}$ between $\mathbf{u}_{i,d}$ and $\mathbf{u}_k$:
\begin{equation}
	\label{eq:dis}
	d_{i,k} = \| \mathbf{u}_{i,d} - \mathbf{u}_k \|.
\end{equation}
The server identifies the nearest cluster $k$ based on these calculated distances and transmits the associated \textit{learnGene} $\Theta_{\mathcal{G}_k}$ from that cluster to the requested agnostic client for model initialization.
For the agnostic client model $\theta_{k,i}$, the initialization parameters consist of two components: inherited cluster \textit{learnGene} $\Theta_{\mathcal{G}_k}$ and the rest randomly initialized $\theta_{0}$, expressed as $\theta_{k,i} = [\theta_{0}; \Theta_{\mathcal{G}_k}]$.
	
	Algorithm~\ref{alg:SERVER} describes the aggregation of \textit{learnGene}, referred to as collaborating learning, as well as the initialization process for the agnostic client model.
	\begin{algorithm}[!htb]
		\caption{GENE-FL: Collaboration and Initialization}
		\label{alg:SERVER}
		\begin{algorithmic}[1]% 添加行号
			\STATE{\bfseries Input:}
			$N_k$ participants in the $k$-th cluster, previous cluster model $\tilde{\Theta}_k$.
			\STATE{\bfseries Output:} Cluster \textit{learnGene} $\Theta_{\mathcal{G}_k}$.
			\STATE \textbf{Server Execution:}
			\STATE Obtain local \textit{learnGene} $\{\theta_{\mathcal{G}_{k,i}}\}_{i=1}^{N_k}$ $\triangleright$ \textbf{Algorithm~\ref{alg:dflg}}
			\STATE Compute $\Theta_{\mathcal{G}_k^{(l)}} \leftarrow \frac{1}{N^{(l)}_k} \sum_{i=1}^{N^{(l)}_k} \theta_{\mathcal{G}_{k,i}^{(l)}}$ using Eq.~(\ref{eq:agg1})
			\STATE Update cluster model $\Theta_k \leftarrow [\Theta_{\mathcal{G}_k}; \tilde{\Theta}_k]$
			\STATE Send $\Theta_k$ to participating clients
			\IF{Agnostic client request}
			\STATE Select the nearest cluster $k$ using Eq.~(\ref{eq:dis})
			\STATE Send the cluster \textit{learnGene} $\Theta_{\mathcal{G}_k}$
			\STATE Update $\mathbf{u}_{k} \leftarrow \frac{N_{k} \cdot \mathbf{u}_{k} + \mathbf{u}_{i,d}}{N_{k} + 1}$
			\ENDIF
			\STATE \textbf{Agnostic Client Initialization:}
			\STATE Send $\mathbf{u}_{i,d}$ using Eq.~(\ref{eq:svd}) to server
			\STATE Receive $\Theta_{\mathcal{G}_k}$ from server
			\STATE Initialize $\theta_{k,i} \leftarrow [\theta_{0}; \Theta_{\mathcal{G}_k}]$
		\end{algorithmic}
	\end{algorithm}
	
	\subsection{Privacy Analysis}\label{sec:priany}
	Typically, the initialization of agnostic client models benefits from the server-side model, while collaborative learning among clients necessitates communication with the server. Therefore, the proposed method should emphasize the importance of privacy guarantees on the server side.
	In the validation phase of this study, we treat the server as a malicious entity capable of reconstructing the original data from a victim client using the iDLG method~\cite{zhu2019deep,wu2021fedcg}. The $\mathcal{L}_{D}$ loss associated with recovering the true data from the victim client $i$ is calculated as follows:
	\begin{equation}
		\mathcal{L}_{D} = \|\nabla_{\Theta} \mathcal{L}_{cls}(x_i) - \nabla_{\Theta} \mathcal{L}_{cls}(\tilde{x}) \|^2,
	\end{equation}
	where $x_i$ is the real data of victim client $i$ while $\tilde{x}$ is the variable to be trained to approximate $x_i$ by minimizing $\mathcal{L}_{D}$ that is the distance between $\nabla_{\Theta} \mathcal{L}_{cls}(x_i)$ and $\nabla_{\Theta} \mathcal{L}_{cls}(\tilde{x})$. The former is observed gradients of $\mathcal{L}_{cls}$ (see Eq.~(\ref{eq:cls})) w.r.t. model parameters $\Theta$ for the real data $x_i$, while the latter is estimated gradients for $\tilde{x}$. 
	We evaluated the privacy guarantees of the GENE-FL, FEDAVG~\cite{mcmahan2017communication}, and PARFED~\cite{sun2021partialfed} methods. For the FEDAVG, which shares the entire network, we set $\Theta := \theta_i$. For PARFED, where only selected network layers are uploaded to the server, $\Theta := [\theta_{0}; \theta_s]$ that $\theta_{0}$ is the random initialization parameter. Similarly, in GENE-FL, only the \textit{learnGene} is shared, so $\Theta := [\theta_{0}; \theta_{\mathcal{G}_i}]$.
	Since the network used for training is ResNet model, we employ the same network for validation, with MSE utilized as the loss function to evaluate the quality of the image reconstruction.
	\section{Experiments}
	\subsection{Experimental Setup}
	\subsubsection{Datasets and Data partition}
Our experiments are conducted on the following three real-world datasets: SVHN~\cite{netzer2011reading}, CIFAR-10~\cite{krizhevsky2009learning}, and CIFAR-100~\cite{krizhevsky2009learning}. SVHN is a benchmark digit classification dataset consisting of 600,000 32$\times$32 RGB printed digit images cropped from Street View house numbers. We select a subset of 33,402 images for training and 13,068 images for testing. The CIFAR-10 dataset consists of 60,000 32$\times$32 color images across 10 classes, with 6,000 images per class. It consists of 50,000 training images and 10,000 test images. Similarly, the CIFAR-100 dataset contains 100 classes of 600 images each, divided into 500 training images and 100 test images per class.

	 To simulate the non-iid distribution in the application scenario, we adopt two strategies for data partitioning: (1) \textit{Sharding}: using the $s$ to control the heterogeneity cross clients and the class distribution of each client is different. For SVHN and CIFAR-10, we choose $s = \{4, 5\}$, while for CIFAR-100, we set it to $s = \{10, 20\}$.
	(2) Dirichlet Distribution Allocation (\textit{DDA}): using the Dirichlet distribution Dir($\beta$) with $\beta$ = \{0.1, 0.5\} on three datasets, creating different local data size and label distributions for all clients. The agnostic clients and known clients are taken from the same dataset, but their classes have no intersection and there is no overlap between samples. 
	\begin{table*}[!htbp]
		\caption{Comparison with state-of-the-art methods on \textit{Comm} (GB, $\downarrow$) and Acc (\%, $\uparrow$) metrics during the known clients updating and \textit{learnGene} condensation under the \textit{Sharding} strategy. Note that we highlight the \textbf{Best} results in bold and the \underline{Second-best} results are underlined.}
		\vskip -0.1in
		\label{comm}
		\begin{center}
			\begin{small}
				\begin{sc}
					\scalebox{0.92}{\begin{tabular*}{20cm}{@{\extracolsep{\fill}}llrcrcrcrcrcrc}
							\toprule[1.2pt] 
							\label{comm}
							% \columnseprulecolor{[HTML]{C0C0C0}}
							&& \multicolumn{4}{c}{\textbf{CIFAR-10}}  & \multicolumn{4}{c}{\textbf{\textbf{CIFAR-100}}} & \multicolumn{4}{c}{\textbf{SVHN}} \\
							\cmidrule(r){3-6}\cmidrule(r){7-10}\cmidrule(r){11-14}
							&\textbf{\# Comm.}	& \multicolumn{2}{c}{$s$ = 4} & \multicolumn{2}{c}{$s$ = 5} & \multicolumn{2}{c}{$s$ = 10} & \multicolumn{2}{c}{$s$ = 20} & \multicolumn{2}{c}{$s$ = 4} & \multicolumn{2}{c}{$s$ = 5} \\
							\cmidrule(r){3-4}\cmidrule(r){5-6}\cmidrule(r){7-8}\cmidrule(r){9-10}\cmidrule(r){11-12}\cmidrule(r){13-14}
							\multirow{-3}{*}{\textbf{Methods}}&\textbf{Params}& \textit{Comm} & Acc &  \textit{Comm} & Acc &  \textit{Comm} & Acc &  \textit{Comm} & Acc &  \textit{Comm} & Acc &  \textit{Comm} & Acc \\
							\midrule
							FEDAVG & \textit{Full model}  & \textit{41.66} & 60.89 & \textit{41.66} & 58.99 & \textit{41.83} & 21.68 & \textit{41.83} & 21.29& \textit{41.66} & 72.03 & \textit{41.66} & 76.93 \\
							FEDBN   & \textit{N-Batch norm.}   & \textit{41.60} &  \underline{70.93} & \textit{41.60 }&  66.90 & \textit{41.78} & 40.22 & \textit{41.78} & 28.60& \textit{41.60} & 82.17 & \textit{41.60} & 78.23 \\
							PARFED &\textit{Partial model} & \textit{\underline{11.67}} & 68.13 & \textit{\underline{11.67}} & 63.23 & \textit{\underline{11.67}} & \underline{45.20} & \textit{\underline{11.67}} & 32.53& \textit{\underline{11.67}} & 83.24 & \textit{\underline{11.67}} & 81.20 \\
							FEDFINA   & \textit{Partial model}  & \textit{30.76} & 65.12 & \textit{30.76} & 61.06 & \textit{30.76} & 40.72 & \textit{30.76} & 31.10 & \textit{30.76} & 78.01 & \textit{30.76} & 74.76\\
							FEDLP   & \textit{Pruning model}  & \textit{27.22} & 62.16 & \textit{16.24} & \underline{67.68} &\textit{13.18} & 42.76& \textit{13.35} & \textbf{33.48}&\textit{36.31} & \underline{86.73} & \textit{13.96} & \underline{82.62}  \\
							
							FEDLPS   & \textit{Pruning model} &  \textit{16.11} & 41.19 & \textit{13.72} & 36.90 & \textit{13.72 }& 40.63 & \textit{13.72} & 28.37&  \textit{13.72} & 83.77 & \textit{13.72} & 75.77\\
							FEDLG & \textit{learnGene} & \textit{17.85} & 66.05 & \textit{17.85} & 64.91 & \textit{17.85} & 41.63 & \textit{17.85} & 32.07& \textit{17.85}& 83.71 & \textit{17.85} & 78.45 \\
							% FedProto   & \textbf{0.11}  & 78.87 & \textbf{0.13}  & 72.06 & \textbf{0.11}  & 69.61 & \textbf{0.13}  & 63.31 & 0.30  & 47.81 & 0.60  & 32.21 \\
							\midrule
							GENE-FL     &\textit{learnGene}  & \textit{\textbf{8.78}}  & \textbf{71.45} & \textit{\textbf{10.33}}&\textbf{ 71.22} & \textit{\textbf{10.26}} & \textbf{46.84} & \textit{\textbf{10.45}} & \underline{33.35}& \textit{\textbf{11.19}} & \textbf{87.30}& \textit{\textbf{11.22}} & \textbf{83.64} \\
							\bottomrule[1.2pt] 
					\end{tabular*}}
				\end{sc}
			\end{small}
		\end{center}
	\end{table*}
	
		\begin{table*}[!htbp]
		\caption{Comparison with state-of-the-art methods on \textit{Comm} (GB, $\downarrow$) and Acc (\%, $\uparrow$) metrics during the known clients updating and \textit{learnGene} condensation under the \textit{DDA} strategy.}
		\vskip -0.1in
		\label{dir-comm}
		\begin{center}
			\begin{small}
				\begin{sc}
					\scalebox{0.91}{\begin{tabular*}{20cm}{@{\extracolsep{\fill}}llrcrcrcrcrcrc}
							\toprule[1.2pt] 
							\label{comm}
							% \columnseprulecolor{[HTML]{C0C0C0}}
							&& \multicolumn{4}{c}{\textbf{CIFAR-10}}  & \multicolumn{4}{c}{\textbf{\textbf{CIFAR-100}}} & \multicolumn{4}{c}{\textbf{SVHN}} \\
							\cmidrule(r){3-6}\cmidrule(r){7-10}\cmidrule(r){11-14}
							&\textbf{\# Comm.}	& \multicolumn{2}{c}{$\beta$ = 0.1} & \multicolumn{2}{c}{$\beta$ = 0.5} & \multicolumn{2}{c}{$\beta$ = 0.1}& \multicolumn{2}{c}{$\beta$ = 0.5} & \multicolumn{2}{c}{$\beta$ = 0.1} & \multicolumn{2}{c}{$\beta$ = 0.5} \\
							\cmidrule(r){3-4}\cmidrule(r){5-6}\cmidrule(r){7-8}\cmidrule(r){9-10}\cmidrule(r){11-12}\cmidrule(r){13-14}
							\multirow{-3}{*}{\textbf{Methods}}&\textbf{Params}& \textit{Comm} & Acc &  \textit{Comm} & Acc &  \textit{Comm} & Acc &  \textit{Comm} & Acc &  \textit{Comm} & Acc &  \textit{Comm} & Acc \\
							\midrule
							FEDAVG  &\textit{Full model}    & \textit{41.66} & 45.69 & \textit{41.66} & 32.15 & \textit{41.83} & 12.02 & \textit{41.83} & 13.23 & \textit{41.66} & 40.25 & \textit{41.66} & 45.78\\
							FEDBN  & \textit{N-Batch norm.} & \textit{41.60} & 50.92 & \textit{41.60} & 48.47 & \textit{41.78} & 18.44 & \textit{41.78} & 16.40 & \textit{41.60} & 53.95 & \textit{41.60} & 61.15 \\
							PARFED &\textit{Partial model}  & \underline{\textit{11.67}} & \underline{63.76} & \textbf{\textit{11.67}} & 53.12 & \underline{\textit{11.67}} & 22.49 & \textbf{\textit{11.67}} & 17.04 & \textbf{\textit{11.67}} & \underline{62.26} & \underline{\textit{11.67}} & 42.16\\
							FEDFINA  &\textit{Partial model}   & \textit{30.76} & 60.35 & \textit{30.76} & 48.44 & \textit{30.76} & \textbf{23.20} & \textit{30.76} & 17.16 & \textit{30.76} & 59.98 & \textit{30.76} & 38.81\\
							FEDLP   & \textit{Pruning model}     & \textit{33.67} & 57.46 & \textit{32.01} & \textbf{66.45} & \textit{24.74} & 17.07 & \textit{24.05} & \underline{20.74}& \textit{37.09} & 60.74 & \textit{36.18} & \textbf{71.81} \\
							FEDLPS   & \textit{Pruning model}   & \textit{16.11} & 48.25 & \textit{16.11} & 35.40 & \textit{16.18} & 13.02 & \textit{16.18} & 16.58 & \textit{13.06} & 52.88 & \textit{16.11} & 29.86 \\
							FEDLG & \textit{learnGene} & \textit{17.85} & 62.25 & \textit{17.85} & 51.95 & \textit{17.85} & 22.49 & \textit{17.85} & 15.48& \textit{17.85} & 59.28 & \textit{17.85} & 51.19  \\
							\midrule
							GENE-FL  &\textit{learnGene}   & \textbf{\textit{11.54}} & \textbf{64.85} & \underline{\textit{15.73}} & \underline{65.61} & \textbf{\textit{10.73}}  & \underline{22.74} & \underline{\textit{12.45}} & \textbf{22.46}& \underline{\textit{12.51}} & \textbf{63.53} & \textbf{\textit{10.54}}  & \underline{70.79}\\
							\bottomrule[1.2pt] 
					\end{tabular*}}
				\end{sc}
			\end{small}
		\end{center}
	\end{table*}
	\subsubsection{Baselines}
	To ensure a fair comparison, we selected a series of federated learning methods for comparison, including FEDAVG~\cite{mcmahan2017communication}, which involves the interaction of all model parameters between the server and clients;
	FEDBN~\cite{li2021fedbn}, which involves the use of local batch normalization before averaging the model to mitigate feature offset;
	PARFED~\cite{sun2021partialfed}, which deliberately skips the four convolutional layers of the model; FEDFINA, which incorporates rich information in the last four convolution layers of the model; FEDLPS~\cite{jia2024fedlps} and FEDLP~\cite{zhu2023fedlp}, which use model pruning to compress and reduce communication costs in federated learning algorithms. FEDLG~\cite{wang2023learngene}, a lightweight \textit{learnGene} presented for the first time, extracts information from gradient updates. Clustering is a preprocessing procedure used to mitigate data heterogeneity. To ensure fairness in comparison, all compared methods undergo clustering operations and are implemented under identical data and equipment conditions.

	\subsubsection{Implementation}
	For local model training, we employ the ResNet18 model~\cite{He2016Deep} for performing classification tasks. The optimizer is SGD, configured with a momentum of 0.9 and a learning rate of 0.01. We set the number of clusters ($K$) to 4, the total number of known clients ($N$) to 50, and the number of agnostic clients ($M$) to 50. Both seen and unseen classes are equally distributed, each comprising 50\% of the total number of classes. During the initialization process, only agnostic clients are involved in both training and inference, aiming to validate the effectiveness of the initialization of the inherited model.
	
	The preprocessing phase for client collaborative training and learning gene compression is set to 100 rounds, the unknown client initialization training phase is set to 50 rounds, and the local update phase is set to 10 rounds. The batch size is set to 64.
	We applied K-Means, KNN, and hierarchical clustering algorithms and observed that they exhibited similar performance trends across various FL methods. Therefore, we opted to use the classic K-Means algorithm.
	Following to the hyperparameter settings in the literature, we set the model pruning probability to 0.5 for the FEDLP~\cite{zhu2023fedlp} method and the local model pruning ratio to 0.8 for the FEDLPS~\cite{jia2024fedlps} method.
	We set the threshold $\varepsilon$ to 0.5 for determining the values.
	The higher $\gamma$ the number of \textit{learnGene} layers, the more layers are selected and the better the performance. 
	In the collaborative accumulation process, we select 10 known clients in each training round. During initialization, we select 10 agnostic clients in each round for model initialization, simulating a dynamic federated learning scenario.
	The experiments are conducted on the server equipped with 1 NVIDIA RTX 3090Ti GPU. Each experiment is repeated three times to compute average metrics.
	%We set the batch size to 64, the number of epochs for global collaborative accumulation training to 100, the number of local epochs to 10, and the number of subsequent training epochs for the initialization-agnostic client model to 50. The specific hyperparameters are described in Appendix~\ref{sec:setting}.
	
	\subsection{Experiment Results}
	We conducted a comprehensive evaluation of the GENE-FL, covering three main aspects: communication costs, testing performance of agnostic models, and privacy guarantees. Furthermore, ablation analysis is performed on various loss functions to determine the importance of each component.
	
	\textbf{GENE-FL improves communication efficiency based on \textit{learnGene}.} In FL, communication cost is a critical bottleneck due to the limited upload bandwidth of edge devices, which is typically constrained to 1 MB/s or less. Table~\ref{comm} presents a comparison of the communication cost (\textit{Comm}) required for uploads and model performance (Acc) between the proposed method and state-of-the-art FL approaches under the \textit{Sharding} strategy. Notably, the communication cost for uploads from edge devices is defined as $R \times B \times | W | \times 2$, where \( R \) represents the total number of communication rounds, which in this experiment is set to 100 rounds for training the known client models; \( B \) denotes the bit-width based on single-precision floating point format, and \( | W | \) refers to the total number of model parameters uploaded by each client to the server per round.
	
	Our method requires uploading only the server-side \textit{learnGene} for collaborative learning, which reduces the communication cost by about 32.88 GB of parameters on CIFAR-10 with $s=4$ compared to the full model exchange in the classical FEDAVG approach and this without sacrificing model performance. Generally, GENE-FL outperforms other approaches across various datasets in most cases, whether compared to the PARFED method in the interactive \textit{Partial model}. Notably, on the CIFAR-10 dataset, the performance improvement is particularly significant, surpassing the second-ranked FEDLP method by approximately 3.54\%. This demonstrates that our method achieves higher model accuracy with lower communication costs, making it more efficient in terms of resource utilization and well-suited for FL systems facing communication bottlenecks.
	
	Table~\ref{dir-comm} lists the communication parameters between clients and the server, along with the model's test performance under the \textit{DDA} non-iid strategy for various methods. Our proposed approach ensures model performance with reduced communication parameters, outperforming state-of-the-art methods such as FEDLP in most scenarios. In particular, GENE-FL reduces communication costs through interactive \textit{learnGene}, especially in the CIFAR-100 dataset with $\beta = 0.1$. Compared to the PARFED method, GENE-FL achieves a reduction of 3.04 GB in communication cost. However, its test accuracy is slightly lower than that of FEDFINA, which has a communication cost of 30.76 GB. The superior performance of FEDFINA is due to the fact that, for $\beta = 0.1$, each client has fewer classes, and the final layers of the model, which are optimized for classification, benefit from personalization. In contrast, performance decreases when the number of classes per client is larger, as in the case of $\beta = 0.5$. Overall, our Gene-driven method effectively improves communication efficiency, and the stability of the \textit{learnGene} parameters contributes to improved model performance.

	% Other methods download some parameters of the communication aggregation and randomly initialize the rest. GENE-FL is to download the well-trained \textit{Cluster learnGene}.
	\begin{table*}[!tb]
		\caption{Comparison with state-of-the-art methods on the number of model parameters (\textit{Initia}, MB, $\downarrow$) used during the initialization process and the performance (Acc, \%, $\uparrow$) achieved after model convergence under the \textit{Sharding} strategy.}
		\begin{center}
			\begin{small}
				\begin{sc}
			
					\label{agnostic-per}
					\scalebox{0.91}{
						\begin{tabular*}{20cm}{@{\extracolsep{\fill}}llccc|cc|cc}
							\toprule[1.2pt]
							\multirow{2}{*}{\textbf{Methods}} & \textbf{\# Initia.} & \multirow{2}{*}{\textit{Initia}}& \multicolumn{2}{c|}{\textbf{CIFAR-10}} & \multicolumn{2}{c|}{\textbf{CIFAR-100}} & \multicolumn{2}{c}{\textbf{SVHN}} \\
							\cmidrule(r){4-9}
							&\textbf{Params}& & $s = 4$ & $s = 5$ & $s = 10$ & $s = 20$ & $s = 4$ & $s = 5$ \\
							\midrule
							FEDBN   & \textit{Cluster N-Batch norm.} & \textit{42.60 M}  & 61.54 & 57.96 & 44.01 & 32.04 & 88.82 & 89.80\\
							PARFED  & \textit{Cluster partial model} & \textit{11.95 M}  & 61.43 & 61.75 & \underline{47.94} & \underline{35.82}& 81.74 & 89.50 \\
							FEDFINA & \textit{Cluster partial model} & \textit{31.50 M} & 54.94 & 51.81 & 41.62 & 30.94 & 87.81 & 86.84 \\
							FEDLP   & \textit{Cluster full model} & \textit{42.66 M}& \underline{63.46} & \underline{65.37} & 40.69 & 34.47 & \textbf{90.95} & \textbf{92.58}  \\
							FEDLPS  & \textit{Cluster full model} & \textit{42.66 M} & 59.10 & 45.85 & 44.38 & 32.06& 87.78 & 77.32  \\
							FEDLG   & \textit{Cluster learnGene} & \textit{18.28 M}  & 60.34 & 57.83 & 44.04 & 30.62& 88.20 & 89.52 \\
							\midrule
							GENE-FL & \textit{Cluster learnGene} & \textit{$\approx$ 9.04 M} &   \textbf{66.90} & \textbf{67.76} & \textbf{49.84} & \textbf{37.85}&\underline{89.03} & \underline{90.47} \\
							\bottomrule[1pt]
					\end{tabular*}}
				\end{sc}
			\end{small}
		\end{center}
		
	\end{table*}

	\begin{table*}[!tb]
			\caption{Comparison with state-of-the-art methods on the number of model parameters (\textit{Initia}, MB, $\downarrow$) used during the initialization process and the performance (Acc, \%, $\uparrow$) achieved after model convergence under the \textit{DDA} strategy.}
			\begin{center}
				\begin{small}
					\begin{sc}
						\label{agnostic-dir}
						\scalebox{0.91}{
							\begin{tabular*}{20cm}{@{\extracolsep{\fill}}llccc|cc|cc}
								\toprule[1.2pt]
								\multirow{2}{*}{\textbf{Methods}} & \textbf{\# Initia.} & \multirow{2}{*}{\textit{Initia}}& \multicolumn{2}{c|}{\textbf{CIFAR-10}} & \multicolumn{2}{c|}{\textbf{CIFAR-100}} & \multicolumn{2}{c}{\textbf{SVHN}} \\
								\cmidrule(r){4-9}
								&\textbf{Params}&  &$\beta$ = 0.1 & $\beta$ = 0.5&$\beta$ = 0.1 & $\beta$ = 0.5& $\beta$ = 0.1 & $\beta$ = 0.5\\
								\midrule
								%						FEDAVG&cluster Full model&50.96&56.57&52.23&47.39&15.89&14.25\\
								FEDBN     & \textit{Cluster N-Batch norm.} & \textit{42.60 M} & 56.33 & 49.79 & 17.35 & 16.44 & 55.63 & 72.42\\
								PARFED  &\textit{Cluster partial model} & \textit{11.95 M} & 59.62 & 51.29 & 21.77 & 16.31& 67.34 & 65.58 \\
								FEDFINA     &\textit{Cluster partial model} & \textit{31.50 M} & 58.10 & 47.65 & 21.04 & 15.47 & 61.67 & 58.78\\
								FEDLP     & \textit{Cluster full model}  & \textit{42.66 M} & 52.39 & \underline{58.12} & 13.70 & \textbf{22.80}& \underline{69.41} & 70.96 \\
								FEDLPS     & \textit{Cluster full model}& \textit{42.66 M} & \underline{60.37} & 51.96 & \underline{22.49} & 18.60 & 57.40 & \textbf{79.63}\\
								FEDLG&\textit{Cluster learnGene}  & \textit{18.28 M} & 59.05 & 49.65 & 21.02 & 14.56 & 64.87 & 74.64\\
								\midrule
								GENE-FL    &\textit{Cluster learnGene}  & \textit{$\approx$ 9.04 M} & \textbf{63.85} & \textbf{61.44} & \textbf{23.62} & \underline{18.73} & \textbf{69.86} & \underline{77.11}\\
								\bottomrule[1pt]
						\end{tabular*}}
					\end{sc}
				\end{small}
			\end{center}
	\end{table*}
	\textbf{GENE-FL achieves competitive performance by \textit{learnGene} initialization than baselines.}
	To validate the effectiveness of using a small-scale \textit{learnGene} to initialize client models in dynamic scenarios, we set clients with untrained class distributions for model initialization and optimization convergence.
	% Table~\ref{agnostic-per} fairly reports the parameter information (downloaded from the server) used to initialize the agnostic client model for different model pruning methods, as well as the average model performance of the last 10 rounds after local training and collaborative learning reached convergence. Since the FEDLP and FEDLPS model pruning methods both perform personalized pruning on the local model and then upload it to the server for aggregation to obtain the cluster model, they are directly downloaded to initialize the unknown client. 
	Table~\ref{agnostic-per} compares various model pruning methods on initializing unknown client models (downloaded from the server) and reports the average performance during the final 10 rounds of local training and collaborative learning. FEDLP and FEDLPS directly initialize unknown clients using cluster models obtained through personalized pruning and server aggregation. 
	In contrast, other methods use aggregated parameters, while GENE-FL leverages a pre-trained \textit{Cluster LearnGene} (~9.04 MB per cluster, with each cluster having a unique \textit{LearnGene}).
	GENE-FL outperforms other approaches across most configurations on various datasets, achieving particularly notable improvements on the CIFAR-10 dataset with $s$ = 4, where it exceeds the FEDLP state-of-the-art model pruning method FEDLP by 3.44\%. Notably, GENE-FL also demonstrates superior performance compared to FEDBN, which uses a relatively complete model to initialize the agnostic client model, reducing the cost consumption by 33.56MB. This indicates that \textit{learnGene} is cost-effective in initializing agnostic client models.
	
	% Notably, GENE-FL also demonstrates superior performance compared to FEDBN,  which utilizes the relatively full model for initialization of agnostic client models, showing an improvement of 12.31\%. This demonstrates the efficacy of \textit{learnGene} in initializing agnostic client models, exhibiting its scalability and adaptability to unknown and varying class distributions. 
	\begin{figure}[!hbt]
		\begin{center}
			\centerline{\includegraphics[width=0.88\columnwidth]{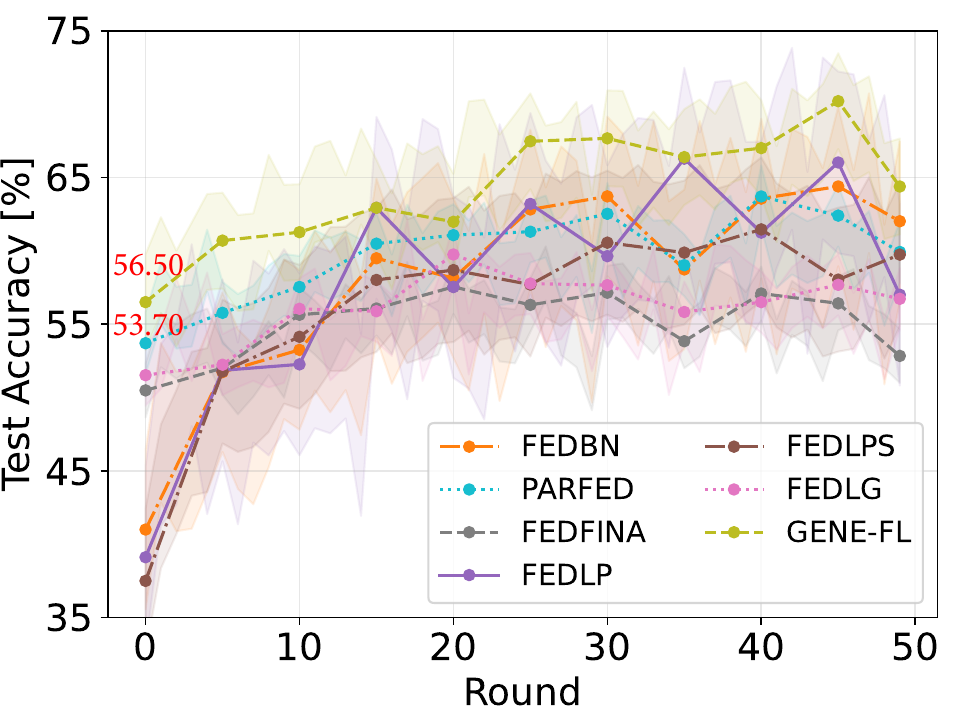}}
			\caption{Performance curve comparison on CIFAR-10 with $s$ = 4. GENE-FL, initialized by the \textit{learnGene}, shows the fastest convergence rate, with its test accuracy increasing rapidly and significantly outperforming other methods.}
			\label{fig:reduce}
		\end{center}
		
	\end{figure}
	
	To further illustrate the importance of model initialization, Figure~\ref{fig:reduce} shows the performance comparison curves of different methods after model initialization when faced with data from untrained class distributions. The results at \textbf{Round 0} represent the average performance after initializing 10 new unknown clients and performing a local update of 10 epochs, with GENE-FL demonstrating superior performance, outperforming the second-best method by 2.80\%. Overall, GENE-FL consistently outperforms others in convergence speed, final performance, and stability, demonstrating superior generalization and scalability. In contrast, FEDLP exhibits greater fluctuations, reflecting performance inconsistencies across training rounds. This highlights the effectiveness of inheriting \textit{learnGene} parameters for initializing agnostic models in dynamic FL scenarios.

	Similar to Table~\ref{agnostic-per}, Table~\ref{agnostic-dir} compares the performance of models trained to convergence after initializing unknown client models under the DDA strategy. The FEDBN method uses parameters from a pre-trained cluster batch norm model downloaded from the server to initialize agnostic client models, while the FEDLP and FEDLPS methods initialize with aggregated cluster models. In contrast, our method initializes the model using cluster \textit{learnGene}. The results show that GENE-FL achieves excellent performance across different dataset settings, especially on CIFAR-10, where it outperforms the FEDBN method by about 8\%. This suggests that inheriting more parameters from pre-trained models is not always beneficial, as it may lead to overfitting on known data and limit adaptation to new data with different class distributions. In addition, models trained using the DDA strategy on non-IID data generally perform worse than those trained using the sharding strategy. This indicates that the former has higher data heterogeneity and greater client-specific class distribution differences, making it more difficult to ensure model performance, reduce costs, and initialize models for unknown clients. Overall, our proposed method outperforms other methods in most scenarios.

	\textbf{GENE-FL guarantees privacy through interactive \textit{learnGene}.}
	We conduct Peak Signal to Noise Ratio (PSNR) as a metric to quantify the similarity between original images and those reconstructed by iDLG~\cite{zhu2019deep,wu2021fedcg}~\footnote{The detailed verification process is in~\ref{sec:priany}.}. A higher PSNR value indicates greater similarity between the images being compared.
	We integrate differential privacy into the FEDAVG by introducing Gaussian noise with noise levels that $\sigma^2$ = 0.001 to the common gradients.
	
\begin{figure}[!tbp]
		\centering
		\includegraphics[width=1\linewidth]{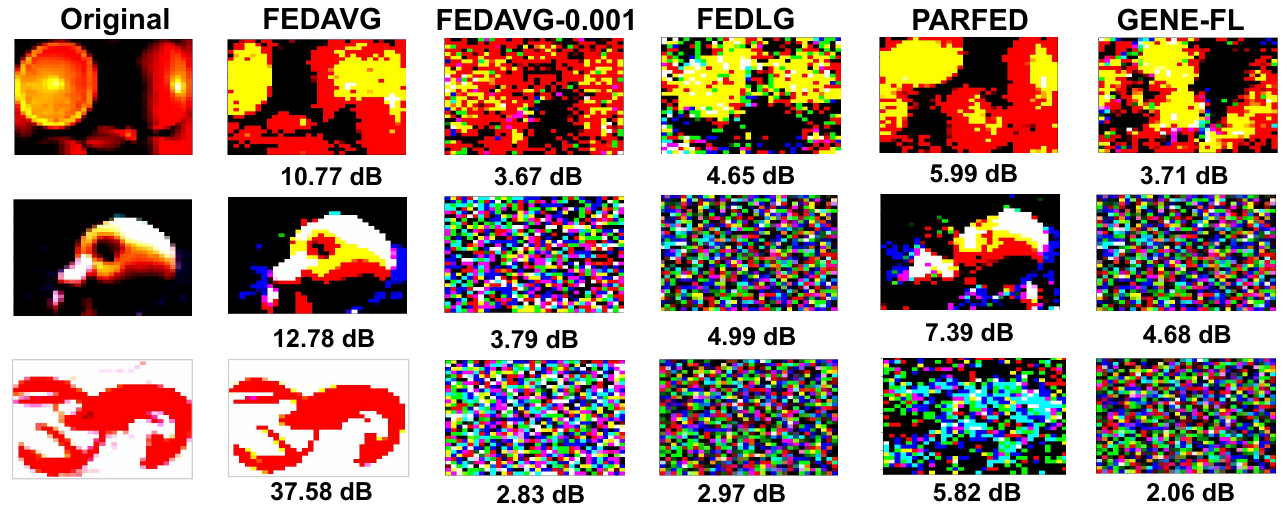}
		\caption{Higher privacy protection. Reconstructing images under iDLG attacks in FEDAVG, FEDLG, PARFED, and the proposed method. Images are extracted from CIFAR-10 and CIFAR-100 datasets, with corresponding PSNR reported beneath each recovered image.}
		\label{priva}
		
	\end{figure}

	Figure~\ref{priva} shows a malicious server attack on client data and subsequent image reconstruction using iDLG across different FL methods with different levels of privacy.
	FEDAVG produces reconstructed images that closely resemble the original, while differential privacy mechanisms show significant improvements. The FEDLG and PARFED methods upload only a subset of model parameters, providing significant privacy benefits. 
	Note that in the last row of images, FEDAVG has a high PSNR value of 37.58dB, closely resembling the original image, while our method renders them indistinguishable from the original in terms of perceptual similarity. This emphasizes that initializing agnostic client models based on the \textit{learnGene} downloaded from the server can prevent privacy leakage, even if the server is malicious.
	
	\textbf{Ablation Study.} To highlight the contribution of each component or our method to the overall performance, we perform a series of ablation experiments. GENE-FL consists of two integral components: 
	\begin{table}[!t]
		\caption{Ablation studies for the GENE-FL.}
		\label{tab:le}
		\begin{center}
			\begin{small}
				\begin{sc}
					\scalebox{0.8}{
						\begin{tabular}{l|ll|cccc}
							\toprule[1.2pt]
							&&  & \textbf{SVHN} & \textbf{CIFAR-10} & \textbf{CIFAR-100}\\
							\multirow{-2}{*}{\textbf{Settings}} &\multirow{-2}{*}{ $\mathcal{L}_{gen}$ } &\multirow{-2}{*}{$\mathcal{L}_{elg}$} & $s$ = 4 & $s$ = 4 & $s$ = 10\\
							\midrule
							GENE-FL w/o $\mathcal{L}_{gen}$  &\XSolidBrush & \Checkmark &  84.78 &  64.69 & 47.25\\
							GENE-FL w/o $\mathcal{L}_{elg}$ & \Checkmark & \XSolidBrush &  87.25 &65.54  &  48.74 \\
							GENE-FL &\Checkmark & \Checkmark& \textbf{89.03} &\textbf{66.90} & \textbf{49.84} \\
							\bottomrule[1.2pt]
					\end{tabular}}
				\end{sc}
			\end{small}
		\end{center}
	\end{table}
	
	(1) The $\mathcal{L}_{gen}$ to learn the knowledge of others with similar clients and find generalizable \textit{learnGene}. (2) The $\mathcal{L}_{elg}$ to focus on learning elastic \textit{learnGene} to improve the knowledge content of the \textit{learnGene}.
	The results in Table~\ref{tab:le} clearly illustrate that both $\mathcal{L}_{elg}$ and $\mathcal{L}_{gen}$ contribute significantly to the performance of the model under various settings. The combined use of both components provides the best results on different datasets, reinforcing the effectiveness of GENE-FL.

	\textbf{Hyperparameter Study.} We exploit different hyperparameters (including $K$, $\varepsilon$, $\lambda_1$, and $\lambda_2$) of proposed method on the CIFAR-10 with $s$ = 4 as shown in Table~\ref{tab:le1}. For the hyperparameter $K$ that controls multiple cluster models on the server side, we observed that multiple global models are more conducive to agnostic clients selecting the optimal initialization model, which verifies the effectiveness of our solution. For the hyperparameter $\varepsilon$, which controls the elastic \textit{learnGene} component in the localization clustering model, a relatively balanced value of 0.5 achieves the best performance. For the hyperparameter $\lambda_1$, which controls the constrained learning based on the clustering model, and the hyperparameter $\lambda_2$, which controls the constrained learning based on the elastic \textit{learnGene}, values of approximately 0.5 and 0.05, respectively, achieve the best performance of the proposed method. These hyperparameter sets are applied to different datasets and consistently achieve good performance.
	\begin{table*}[!htbp]
		\caption{Ablation study on various hyperparameters.}
		\begin{center}
			\begin{small}
				\begin{sc}
					\label{tab:le1}
					\renewcommand\tabcolsep{1.0pt}
					\scalebox{0.88}{
						\renewcommand{\arraystretch}{1.1}
						% 第一部分表格
						\begin{tabular*}{20cm}{@{\extracolsep{\fill}}lccc|ccc|ccc|ccc}
							\toprule[1.2pt]
							& \multicolumn{3}{c|}{$K$} & \multicolumn{3}{c|}{$\varepsilon$}&\multicolumn{3}{c|}{$\lambda_1$}& \multicolumn{3}{c}{$\lambda_2$}\\
							
							& 1 & 4 &10& 0.1 & 0.5 & 0.9 &0.05&0.1&0.5&0.05&0.1&0.5\\
							\midrule
							Acc (\%) & 64.89 & \textbf{66.90} &62.29& 65.17 & \textbf{66.90} & 63.11&64.24&63.49&\textbf{65.54}&\textbf{64.49}&63.06&60.56\\
							\bottomrule[1.2pt]
					\end{tabular*}}
				\end{sc}
			\end{small}
		\end{center}
	\end{table*}
	
	To ensure a fair comparison in the experiments, each of the compared methods underwent a clustering preprocessing procedure, resulting in multiple clustered models. During the initialization process, an unknown client selects the initialization model that best fits its own data distribution. To validate the effectiveness of the clustering preprocessing step, Table~\ref{tab:clu} presents a comparison of the initialization model performance on the CIFAR-100 dataset with high heterogeneity ($s=20$) under two conditions: one without clustering ($k=1$) and one with clustering ($k=4$). The results demonstrate that the clustering preprocessing step is effective, highlighting that in dynamic federated learning (FL) scenarios, selecting an initialization model that aligns with the client’s data distribution is more effective than using a general global model.
	\begin{table*}[!htbp]
		\caption{Ablation study on various hyperparameters.}
		\begin{center}
			\begin{small}
				\begin{sc}
					\label{tab:clu}
					\renewcommand\tabcolsep{1.0pt}
					\scalebox{0.88}{
						\renewcommand{\arraystretch}{1.1}
						% 第一部分表格
						\begin{tabular*}{20cm}{@{\extracolsep{\fill}}lcc|cc|cc|cc|cc|cc}
							\toprule[1.2pt]
							& \multicolumn{2}{c|}{FEDBN} & \multicolumn{2}{c|}{PARFED} & \multicolumn{2}{c|}{FEDFINA}&\multicolumn{2}{c|}{FEDLP}& \multicolumn{2}{c|}{FEDLPS}&\multicolumn{2}{c}{FEDLG}\\
							
							& 1 & 4 &1& 4& 1 & 4 &1&4&1&4&1&4\\
							\midrule
							Acc (\%) &30.97&\textbf{32.04}& 35.40 & \textbf{35.82} &28.35& \textbf{30.94} & 33.61 & \textbf{34.47}&29.23&\textbf{32.06}&30.20&\textbf{30.62}\\
							\bottomrule[1.2pt]
					\end{tabular*}}
				\end{sc}
			\end{small}
		\end{center}
	\end{table*}

	\begin{figure}[!hbt]
		\begin{center}
			\centerline{
				\includegraphics[width=0.9\columnwidth]{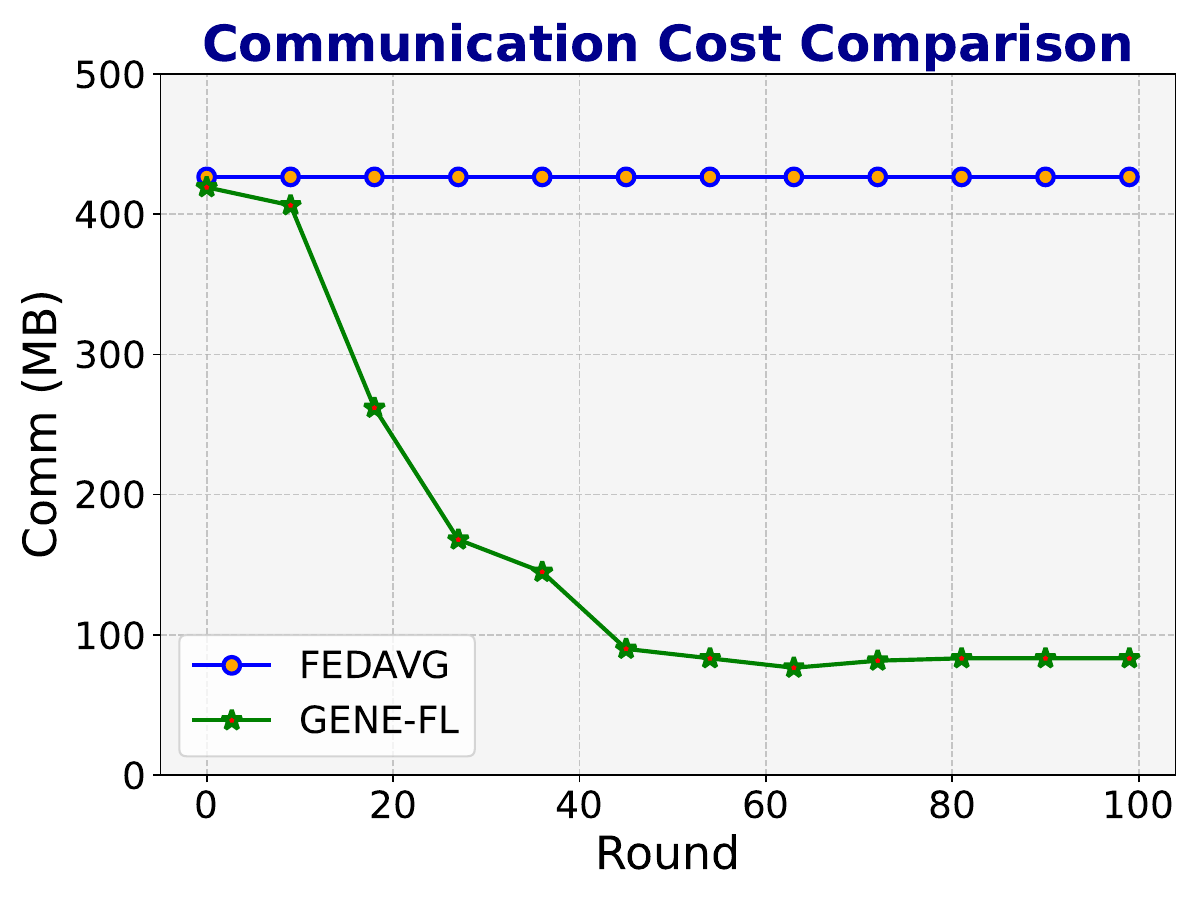}}

			\caption{Comparison of communication cost curves from client to server over communication rounds on the CIFAR-10 dataset under $s$=5 data partition.}
			\label{fig:cost}
		\end{center}
	\end{figure}
	
		\textbf{Communication Cost Analysis.}
	We conducted a comparative analysis of the communication cost curves for the FEDAVG and GENE-FL methods, focusing on the cumulative cost of client uploads to the server, as shown in Figure~\ref{fig:cost}. The FEDAVG method maintains a constant communication cost throughout all rounds since it involves transmitting the complete model parameters in each round. In contrast, GENE-FL shows a significantly reduced and progressively declining communication cost. This trend is attributed to the optimization of \textit{learnGene} during collaborative training, which effectively compresses the transmitted information. Furthermore, the well-trained \textit{learnGene} enable efficient initialization of unknown client models by extracting and inheriting generalized invariant knowledge across models. This not only ensures effective model personalization but also substantially minimizes the overall communication overhead, making GENE-FL more communication efficient compared to FEDAVG.

		\begin{figure}[!hbt]
		\begin{center}
			\centerline{\includegraphics[width=0.9\columnwidth]{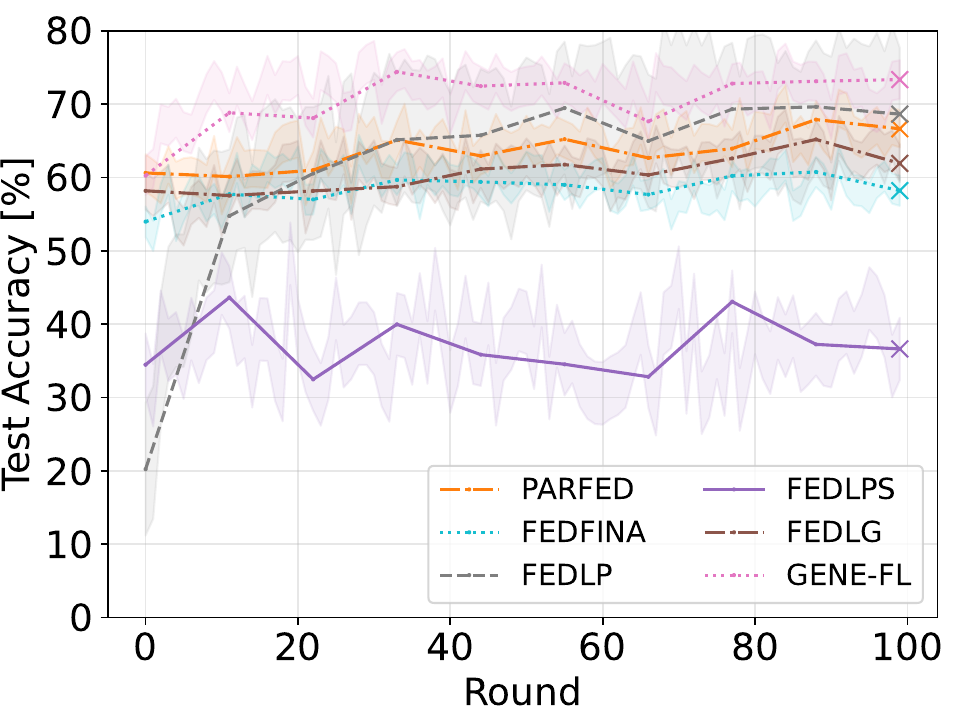}}
			\caption{Illustration of the accuracy progression over communication rounds on the CIFAR-10 dataset under $s$=5 partition.}
			\label{fig:convergence}
		\end{center}
	\end{figure}
	\textbf{Convergence Analysis.} The convergence analysis of GENE-FL, as shown in Figure~\ref{fig:convergence}, demonstrates that the introduction of \textit{learnGene} communication does not lead to convergence issues. We present the accuracy progression of communication rounds on the CIFAR-10 dataset under \textit{Sharding} partitioning. It is observed that the accuracy of the GENE-FL method steadily increases and gradually converges. Unlike local training, which may lead to overfitting and subsequently a decrease in accuracy due to excessive training, our method effectively avoids the overfitting problem. Moreover, compared to baseline methods that risk premature convergence and getting trapped in local optima, GENE-FL consistently seeks better solutions. It achieves optimal performance before final convergence, ensuring superior generalization and performance compared to other methods.

	\section{Conclusions}
		In this paper, we delve into the challenges of high communication interaction costs and model initialization for agnostic clients in dynamic agnostic federated learning. We present a Gene-driven parameter-efficient dynamic Federated Learning framework consisting of three modules: \textit{learnGene} Condensation via Smooth Updating, \textit{learnGene} Collaborative Aggregation, and \textit{learnGene} Initial Agnostic Client Model that effectively addresses the aforementioned challenges. The effectiveness of the proposed approach has been extensively validated on various classification tasks against several popular methods. In the future, we will further investigate how agnostic heterogeneous models can be effectively integrated with Learngene to address initialization and communication issues.

	
	\bibliographystyle{IEEEtran} 
	 \bibliography{example_paper}

\end{document}